\documentclass[a4paper]{jpconf}
\usepackage{graphicx}

\newcommand{\be}{\begin{eqnarray}}
 \newcommand{\ee}{\end{eqnarray}}
 \newcommand{\nee}{\nonumber\end{eqnarray}}
 \newcommand{\nn}{\nonumber\\}

  \newcommand{\bc}{\begin{center}}
 \newcommand{\ec}{\end{center}}
 \newcommand{\pup}{p^\uparrow}

\newcommand{\qup}{q^\uparrow}

\def\kt{k_\perp}
\def\bkt{\bf k_\perp}

\def\avk{\langle k_\perp ^2\rangle}
\def\avp{\langle p_\perp ^2\rangle}
\def\avPT{\langle P_T^2\rangle}
\def\xb{x_{_{\!B}}}

\def\S{_{_S}}
\def\T{_{_T}}
\def\C{_{_C}}

\def\BM{_{_{B\!M}}}

\def\s              {\sigma}

\begin{document}
\title{ Tests for Sivers,  Boer-Mulders and
transversity distributions in difference cross sections in SIDIS }

\author{Ekaterina Christova}

\address{Institute for Nuclear Research and Nuclear Energy, Bulgarian Academy of Sciences, Sofia, Bulgaria}

\ead{echristo@inrne.bas.bg}

\author{Elliot Leader}

\address{Imperial College, London, UK}

\ead{e.leader@imperial.ac.uk}

\begin{abstract}
A major experimental program  is presently underway  to determine
the Sivers, B\"{o}er-Mulders and transversity distributions, vital
for  understanding  the internal structure of the nucleon. To this
end  we consider  the  Sivers, B\"{o}er-Mulders and transversity
azimuthal asymmetries of the difference cross sections of hadrons
with opposite charges in SIDIS  reactions with unpolarized and
transversely polarized target $l+N\to l'+h+X$, $h=\pi^\pm ,K^\pm
,h^\pm$. We show that on deuteron  target these asymmetries are
particularly simple and determine the sum of the valence-quark
$Q_V=u_V+d_V$ transverse momentum dependent distributions  without
any contributions from the  strange or other sea-quark  functions.

At present, data on these asymmetries are presented for the
integrated asymmetries i.e. the $x_B$- and $z_h$-dependent
asymmetries. If data are available in small bins in $Q^2$,  so that
$Q^2$-dependence  can be neglected,
   these expressions simplify dramatically
leading to remarkably simple and powerful tests of the
 simplifying assumptions used in extracting these functions from the data.\end{abstract}

\section{Introduction}

We consider the transverse momentum dependent (TMD) Sivers,
Boer-Mulders (BM) and transversity functions, measured in
transversely polarized or  unpolarized  semi-inclusive deep
inelastic scattering (SIDIS) $l+N\to l'+h+X$.

At present, the extraction of these transverse momentum dependent
(TMD) functions has been relatively simplistic, involving two key
conventions:

a) The analysis is in leading order in perturbative QCD

b) Both the  parton distributions (PDFs) and fragmentation functions
(FFs)
 which depend on parton intrinsic momentum $k_\perp $, are typically parametrized under the following conventions:  1)
a factorized form for the $\xb/z_h$ and $\kt$-dependences:
 \be
 \Delta f(\xb \, {\rm or}\, z_h, k_\perp^2)=\Delta f(\xb \,
{\rm or}\, z_h) \,e^{-k_\perp^2/\avk}\,\cdot
 \ee
 2) the $\xb (z_h)$-dependence is proportional to the collinear PDFs (FFs),
 3) the $Q^2$-evolution is in the collinear PDFs (FFs) and
 4) the $\kt $--dependence is  Gaussian type and  flavour independent.
 The functional form of the TMD functions is
 still under discussion and these are the simplest, commonly used at present {\it standard}
 parametrizations.  Additional assumptions about the sea-quark contributions are made as well.
 For BM functions it is assumed that they are proportional to Sivers functions.

  The goal of our studies is to show that if one uses the so called
  "difference" asymmetries one can
  1) obtain information about the valence-quark TMDs without
  contributions from the sea-quarks, 2) test the standard
  parametrization and  the assumption about BM function using relations between
   measurable quantities only. These tests are crutial as they
   would verify to what extend  the used standard parametrizations provide a reliable information
    on the TMD distributions for the present set of data.

The "difference" cross section asymmetries are combinations of the
type:
 \be
  A^{h-\bar h}\equiv \frac{\Delta \s^h-\Delta \s^{\bar h}}{\s^h-\s^{\bar h}}\label{A}
\ee
 where $\s^h$ and $\Delta\s^h=\s^{h\uparrow}-\s^{h\downarrow}$ are the unpolarized and polarized cross sections respectively.
 The arrows indicate the polarization of the target, and $h$ and $\bar h$ are hadrons with opposite charges.

Previously  the difference cross section asymmetries \cite{we1,we2}
have appeared rather useful in the simple collinear picture and the valence
quark helicity parton densities and fragmentation functions were
determined directly \cite{COMPASS}.
 Here we extend our studies to the  non-collinear picture when transverse momentum dependence is included.

 Presently, data on Sivers, Boer-Mulders (BM) and
  transversity asymmetries are  presented as functions of only one of the
kinematic variables $\xb$ or  $z_h$, sometimes with, in addition, the
$Q^2 $ and  $P_T$ dependence, with integration over the measured
intervals of the other  variables.
 We obtain analytic expressions for the $\xb$ and $z_h$ dependence of the  asymmetries.
These expression strongly simplify  if in the measured kinematic
intervals the binning in $Q^2$ is small enough to allow the neglect
 of the $Q^2$-dependence of the parton densities and  FFs.
 Here we shall  present  the general formulae for Sivers asymmetries only, and the formulae when this simplification is valid for all asymmetries.
 Most of the presented results can be found in more details in ref. \cite{we}.

 We present the asymmetries on deuteron  target, when the asymmetries provide information on the sum of the
 valence-quark TMDs $\Delta f_{Q_V},\, Q_V=u_V+d_V$ and the results
 appear especially simple. The results on proton targets are
 obtained with the simple replacement $\Delta f_{Q_V} \to e_u^2 \Delta f_{u_V} -e_d^2\Delta
 f_{d_V}$ for final pions and $\Delta f_{Q_V}\to e_u^2 \Delta f_{u_V}$ for
 final kaons. The explicit expressions for them are given in ref.\cite{we}.

\section{Sivers distribution functions}

The Sivers distribution function $\Delta^N \! f_{q/S_T}(\xb, \kt)$
appears in the expression for
  the number density of unpolarized quarks $q$ with intrinsic transverse momentum
 ${\bf k}_\perp$ in a
 transversely polarized proton $\pup$ with 3-momentum $\bf P$ and transverse spin $\bf S_T$ \cite{Sivers}:
\be
f_{q/\pup}(\xb, {\bf \kt}) = f_{q/p}(\xb, \kt)
+\frac{1}{2}\,\Delta^N \! f_{q/S_T}(\xb, \kt) \,\,
 {\bf S_T}\cdot ({\bf \hat P}\times {\bf \hat k}_\perp)
\ee
where $f_{q/p}(\xb, \kt)$ are the unpolarized $\xb$ and
$k_\perp$-dependent parton densities and the triple product induces
a definite azimuthal $\sin (\phi_h-\phi_ S)$-dependence, $\phi_h$
and $\phi_S$ are the azimuthal angles of the final hadron and the
spin of the target,
  $x_B=Q^2/2(P.q)$ is the usual Bjorken variable.

  To access the Sivers TMDs one considers the $\sin(\phi_h-\phi_S)$
azimuthal moment $A_{UT}^{\sin(\phi_h-\phi_S)}$  of the  transverse
single-spin  asymmetry in SIDIS \cite{general}.
In accordance with this we define  Sivers asymmetry
for the difference cross section  analogously:
 \be
 A_{UT}^{Siv,{h-\bar h}}(\xb , Q^2,z_h,
P_T^2)&=&\frac{1}{S_T}\, \frac{\int d\phi_h d\phi_S \left[d^6
\s^\uparrow-d^6\s^\downarrow \right]^{ h-\bar
h}\,\sin(\phi_h-\phi_S)} {\int d\phi_h d\phi_S \left[d^6
\s^\uparrow+d^6\s^\downarrow \right]^{h-\bar h}}
\ee
where $d^6\s^{\uparrow,\downarrow} = d\s^{\ell
\,p^{\uparrow,\downarrow } \to \ell^\prime h X}/
(d\xb\,dQ^2\,dz_h\,dP_T^2\,d\phi_h\,d\phi_S) $ stands for the differential
cross section of SIDIS with
 an unpolarized lepton beam on a  transversely polarized target
 in the kinematic region  $P_T \simeq k_\perp \ll Q$  at
  order $(k_\perp /Q)$,  $P_T$ is the transverse momentum of the final hadron  in the
$\gamma*-p$\, c.m. frame,
 and $ z_h$, $Q^2$ and $y$ are the usual measurable SIDIS quantities:
\be
 \quad z_h=\frac {(P.P_h)}{(P.q)},\quad Q^2=-q^2, \quad q=l-l',
\quad
 y=\frac{(P.q)}{(P.l)},\quad  Q^2=2ME \xb y
 \ee
 with $l$ and $l'$, $P$ and $P_h$  the 4-momenta of the initial
and final leptons,  and  hadrons. Throughout the
paper we follow the notation and kinematics of ref. \cite{general}.

In general, only the valence-quark
TMD functions enter the difference cross sections.  When a deuteron  target is used,  a further
simplification occurs -- independently of the final hadrons,  only
one parton density enters -- the sum of the valence quarks:
\be
\Delta f_{Q_V/S_T} (\xb,\kt)\equiv \Delta f_{u_V/S_T} (\xb,\kt)+\Delta f_{d_V/S_T} (\xb,\kt)
\ee

Following the convention for the standard parametrizations, we adopt the following  form
for valence-quark Sivers function:
\be
 \label{MS} \Delta f_{Q_V/S_T} (\xb,\kt)=\Delta
f_{Q_V/S_T} (\xb)\,\sqrt{2e}\, \frac{\kt}{M\S } \;
\frac{e^{-\kt^2/\avk\S }}{\pi\avk} \label{Siv-dist}
 \ee
 where
  \be
   \label{curlyN}
   \Delta f_{Q_V/S_T} (\xb,Q^2)=2\,{\cal N}^{Q_V}_{Siv}(\xb)\;Q_V(\xb,Q^2)
    \ee
   and $Q_V=u_V+d_V$  is the sum of the  collinear valence PDFs,
 and
 \be
  \label{kperpSiv}
 \quad \avk\S  = \frac{\avk  \, M^2\S }{\avk  +  M^2\S }\,\cdot
   \ee

  The  unknowns are  $M\S $ (involved in the definition of $\avk\S$)
 and ${\cal N}^{Q_V}_{Siv}(\xb )$. They are to be determined
 in the integrated $z_h$- and $\xb$- Sivers difference asymmetries, for which we obtain:
\be
A_{UT}^{Siv, h-\bar h}(z_h ) &=& {\cal B}_{Siv}^{h}(z_h)\,\frac{z_h}{\sqrt{\avp+z_h^2\avk\S }},\qquad h=\pi^+,K^+,h^+ \label{AS-zh-general}\\
{\cal B}_{Siv}^{h}(z_h)&=& A_{Siv}\,\frac{\int d\xb\,\int
dQ^2\, \frac{1+(1-y)^2}{Q^4}\,\Delta
f_{Q_V/S_T}(\xb,Q^2)\,D_{u_V}^{h}(z_h,Q^2)}
 {2\,\int d\xb\,\int dQ^2\,\frac{1+(1-y)^2}{Q^4}\,Q_V(\xb,Q^2)\,D_{u_V}^{h}(z_h,Q^2)}
\ee
and
\be
A_{UT}^{Siv,{ h-\bar h}}(\xb )&=& {\cal C}_{Siv}^{h}(\xb)\,{\cal
N}^{Q_V}_{Siv}(\xb ),\qquad h=\pi^+,K^+, h^+\label{AS-xb-general}\\
 {\cal C}_{Siv} ^{h}(\xb)&=& A_{Siv}\frac{\int dQ^2\,\frac{1+(1-y)^2}{Q^4}\,Q_V(\xb,Q^2)\,\int
dz_h \,z_h \left[ D_{u_V}^{h}(z_h, Q^2)\right]/\sqrt{\avPT\S  }}
{\int dQ^2\,\frac{1+(1-y)^2}{Q^4}\,Q_V(\xb,Q^2)\,\int dz_h
\,D_{u_V}^{h}(z_h, Q^2)} \,\cdot
 \ee
Eqs. (\ref{AS-zh-general}) and (\ref{AS-xb-general}) determine ${\cal N}^{Siv}(\xb)$ and
$\avk\S$ without any contributions from the sea quarks.

For bins corresponding to a reasonably small interval $\Delta Q^2$ in $Q^2$, we replace the integral
over $Q^2$ by $\Delta Q^2 $ times the $Q^2$-dependent functions evaluated at the mean value $\bar {Q}^2$ for the bin.
 Then  (\ref{AS-zh-general}) and (\ref{AS-xb-general})
become particularly simple:
\be
 A_{UT}^{Siv, h-\bar h}(z_h,\bar {Q^2} )&=&
 {\bar B}_{Siv}(\bar {Q^2})\,\frac{z_h}{\sqrt{\avp +z_h^2 \avk\S }},\label{ASivzd}\\
A_{UT}^{Siv,{ h-\bar h}}(\xb,\bar {Q^2} )&=& \bar C_{Siv}^{h}
(\bar{Q^2})\,{\cal N}^{Q_V}_{Siv}(\xb )\,,\label{xb_S_approx}
 \ee
as $ {\bar B}_{Siv}$ and $\bar C_{Siv}^{h}$ are constant factors:
\be
 {\bar B}_{Siv}(\bar
{Q^2})&=& A_{Siv}\,\frac{\int d\xb\,\,[1+(1-\bar{y})^2] \, \Delta f_{Q_V/S_T}(\xb,\bar{Q^2})}
{2\,\int d\xb\,\,[1+(1-\bar{y})^2]\,Q_V(\xb,\bar
{Q^2})}, \quad \forall\, h\label{BSivzd}\\
 \bar C_{Siv}^{h}(\bar{Q^2})&=& A_{Siv}\,\frac{\int dz_h \,z_h
\left[ D_{u_V}^{h}(z_h,\bar {Q^2})\right]/\sqrt{\avPT\S  }} {\int dz_h \,D_{u_V}^{h}(z_h,\bar {Q^2})},\quad
h=\pi^+, K^+\label{CSivxd}
 \ee
 Here
 \be
 \bar{y} = \frac{\bar{Q}^2}{2MEx_B},\quad A_{Siv}=\frac{\sqrt{e\pi}}{2\sqrt{2}}\,\frac{\avk\S  ^2 }{M\S \,\avk},\quad \avPT\S
=\avp +z_h^2\avk\S
 \ee
 Eq. (\ref{ASivzd}) is remarkably strong prediction both for its explicit $z_h$ behaviour, determined solely by the  Gaussian
 dependence on $\kt$,  and for its
 independence of $h$.
 Eq. (\ref{BSivzd}) implies that Sivers $\xb$-asymmetries  should have the same $\xb$-behaviour for all final hadrons.

  %%%%%%%%%%%%%%%%%%%%%%%%%%%%%%%%%%%%%%%%%%%%%%%%%%%%%%%%%%%%%%%%%%%%%%%%%%%%%%%%%%%%%%%%%%%%%%%%%%%%%%%%%%%%%%
\section{Boer-Mulders distributions}
The
extraction of Boer-Mulders and transversity distributions is
more complicated as compared to Sivers parton densities. The reason
is that Sivers functions enter the cross section in convolution with
the unpolarized TMD fragmentation functions, \emph{known} from
multiplicities, while the BM and transversity functions enter the cross
sections in convolution with the transversely polarized TMD FFs, the
so called Collins functions $\Delta^N D_{h/q\uparrow}(z_h,p_\perp)$. The latter can, in principle,  be extracted from
$e^+e^-\to h_1h_2X$, but at present are rather poorly known.

The Boer-Mulders   function determines the distribution of transversely polarized quarks  $\qup$ in an unpolarized proton $p$
  \cite{BM}:
\be
 \Delta^N
\! f_{\qup/p}(x_B, \kt) \equiv \Delta f^{q}_{s_y/p}(x_B,\kt)  = - \frac
{\kt}{m_p} \, h_{1}^\perp (x_B, \kt)\,\cdot \label{b-m}
\ee
It is accessed  measuring the $\cos 2\phi_h$-momentum  in unpolarized SIDIS differential cross section.
The difference BM asymmetry is defined respectively:
\be
A\BM^{h-\bar h}=\frac{\int d\phi_h \,\cos
2\phi_h\,d^5\s^{h-\bar h}}{\int d\phi_h\,d^5\s^{h-\bar h}}\,\cdot
\ee
On deuteron  target, independently of the final  hadrons $h-\bar h$, it determines only
the sum of the valence quark BM  distribution:
\be
   \Delta f^{Q_V}_{s_y/p}(\xb ,k_\perp )\equiv \Delta f^{u_V}_{s_y/p}(\xb ,k_\perp )+ \Delta f^{d_V}_{s_y/p}(\xb ,k_\perp )
   \ee
that is parametrized accordingly:
\be
\hspace*{-.5cm}  \Delta  f^{Q_V}_{s_y/p}(\xb,\kt ,Q^2) \!&=&\! \Delta  f^{Q_V}_{s_y/p}(\xb,Q^2)\;
\sqrt{2e}\,\frac{\kt}{M\BM} \; \frac{e^{-\kt^2/\avk_{BM}
}}{\pi\avk_{BM}}\nn
\hspace*{-.5cm} \Delta  f^{Q_V}_{s_y/p}(\xb,Q^2)\!&=&\! 2\,{\cal
N}\BM^{Q_V}(\xb)\,Q_V(\xb,Q^2), \label{BM-Vdist}
 \ee
 where ${\cal N}\BM^{Q_V}(\xb)$ and $M\BM$ are to be determined from the integrated BM asymmetries.
 Here we present only the $\xb$-dependent asymmetry. For all final $h$ it determines the same ${\cal N}^{Q_V}\BM$:
 \be
 A\BM(\xb )^{h-\bar h}= {\cal
C}\BM ^{h}(\xb)\, {\cal N}^{Q_V}\BM (\xb ),\qquad \forall \,h
\label{cos2phi}
\ee
 where for $h=\pi^+, K^+$ we have:
 \be
  {\cal
C}^{h}\BM (\xb)&=&\!\!-4e\,A\BM\,A_{Coll} \,\frac{\,\int\! \int dQ^2\,
 dz_h\,\frac{1-y}{Q^4} \,Q_V(\xb)\,z_h\,
 \Delta^ND_{h/u_V\uparrow}(z_h)\,/\langle P_T^2\rangle\BM }
{\int \int dQ^2\,dz_h
\,\frac{1+(1-y)^2}{Q^4}\,Q_V(\xb)\,D_{u_V}^{h}(z_h)}\,\cdot
 \ee
For $h=h^\pm$ the relative expression is given in \cite{we}.

 If $Q^2$-evolution can be neglected, then  $ A\BM(\xb )^{h-\bar h}$ is determined solely by
 ${\cal N}^{Q_V}\BM$:
\be
\hspace*{-.5cm}A\BM^{h-\bar
h}(\xb,\bar Q^2 )= \frac{1-\bar{y}}{1 + (1-\bar{y})^2}\,\,\bar C^{h}\BM(\bar
Q^2)\,
 {\cal N}^{Q_V}\BM (\xb ),\quad h=\pi^+, K^+, h^+
\ee
where $\bar C^{h}\BM $ are constant factors. This implies, in particular,  that $ A\BM(\xb )^{h-\bar h}$ have the same $\xb$-behaviour:
\be
 A\BM^{ \pi^+-\pi^-}(\xb )\simeq A\BM^{ K^+-K^-}(\xb )\simeq A\BM^{ h^+-h^-}(\xb )
 \quad {\rm or}\quad
 \frac{A_{BM}^{\pi^+-\pi^-}(\xb)}{ A_{BM}^{ h^+-h^-}(\xb)}=.. .={\rm const}
 \ee

%%%%%%%%%%%%%%%%%%%%%%%%%%%%%%%%%%%%%%%%%%%%%%%%%%%%%%%%%%%%%%%%%%%%%%%%%%%%%%%%%%%%%%%%%%%%%%%%%%%%%%%%%%%%%%%%%%%%%%%%%%%%%%%%
\section{Relations between BM and Sivers asymmetries}\label{Relations 1}

 In current analysis an additional simplifying assumption is made, namely
the BM function is assumed proportional  to
its chiral-even partner -- the Sivers function \cite{BM_1}.  In our case this reads:
\be
 \Delta
f_{s_y/p}^{Q_V}(x,k_\perp)=\frac{\lambda_{Q_V}}{2}\,\Delta
f_{Q_V/S_T}(x,k_\perp),\label{BM1}
 \ee
 where $\lambda_{Q_V}$ is a fitting parameter. This implies that BM and Sivers $\xb$-asymmetries measure the same ${\cal N}_{Siv}^{Q_V}$
 and they are related:
 \be
 \frac{A\BM^{h-\bar h}(\xb )}{A_{UT}^{Siv,h-\bar h}(\xb)}
  =
  \frac{\lambda_{Q_V}}{2}\,{\cal C}^h(\xb),
  \quad h=\pi^+, K^+, h^+
  \ee
where ${\cal C}^h (\xb)$ is independent of ${\cal N}_{Siv}$:
  \be
{\cal C}^h(\xb) \!=\!\frac{- 4\sqrt {2e}}{\sqrt \pi}\,
\frac{\int\int
dQ^2\,dz_h\,\frac{1-y}{Q^4}\,Q_V(\xb,Q^2)z_h\,\Delta^N
D_{h/u_V\uparrow}(z_h, Q^2)/\avPT_{\widetilde{BM}} } {\int\int
dQ^2\,dz_h\,\frac{1+(1-y)^2}{Q^4}\,Q_V(\xb,Q^2)z_h\,D_{u_V}^{h}(z_h, Q^2)/\sqrt{\avPT_{Siv}}},\quad h=\pi^+, K^+
\label{R1test}
\ee

If $Q^2$-evolution can be neglected the ratio of the asymmetries is completely fixed:
 \be
 \label{R1simpletest}
 \frac{A\BM^{h-\bar h}(\xb )}{A_{UT}^{Siv,h-\bar h}(\xb)}=
  \frac{\lambda_{Q_V}}{2}\,\frac{1-\bar{y}}{1 + (1-\bar{y})^2}\,\,\bar C^h(\bar Q^2),
   \ee
  where  $\bar C^h$ is a constant factor.

Thus, if data exist for a range of $x_B$ values at the same $\bar{Q}$, eq.~(\ref{R1simpletest})
   should allow a  test of the used connection between BM and Sivers functions without requiring knowledge of
   BM,  Sivers or even Collins functions,  involving only measurable quantities.

   %%%%%%%%%%%%%%%%%%%%%%%%%%%%%%%%%%%%%%%%%%%%%%%%%%%%%%%%%%%%%%%%%%%%%%%%%%%%%%%%%%%%%%%%%%%%%%%%%%%%%%%%%%%%%%%%
%%%%%%%%%%%%%%%%%%%%%%%%%%%%%%%%%%%%%%%%%%%%%%%%%%%%%%%%%%%%%%%%%%%%%%%%%%%%%%%%%%%%%%%%%%%%%%%%%%%%%%%%%%%
\section{Relations between BM and Collins asymmetries}\label{Relations 2}
%%%%%%%%%%%%%%%%%%%%%%%%%%%%%%%%%%%%%%%%%%%%%%%%%%%%%%%%%%%%%%%%%%%%%%%%%%%%%%%%%%%%%%%%%%%%%%%%%%%%%%%%%%%5

 The distribution of transversely polarized quarks $\qup$ in a transversely polarized proton $\pup$ defines the transversity distributions $h_{1q}(\xb)$
 or $ \Delta _T q(\xb)$:
\be
 \Delta _T q(\xb) =  h_{1q}(\xb) = \int \!d^2 {\bkt} \,
h_{1q}(\xb,\kt)\label{int-transv} \,,
\ee
 where $h_{1q}(\xb,\kt)$ is the
transversity distribution depending  on the  parton transverse
momentum. It is selected by the Collins asymmetry, which for the difference cross sections is:
\be
A_{UT}^{\sin(\phi_h+\phi_S),{h-\bar
h}}&=&\frac{1}{S_T}\, \frac{\int d\phi_h d\phi_S \left[d^6
\s^\uparrow-d^6\s^\downarrow \right]^{ h-\bar
h}\,\sin(\phi_h+\phi_S)} {\int d\phi_h d\phi_S \left[d^6
\s^\uparrow+d^6\s^\downarrow \right]^{h-\bar h}}\,\cdot
 \ee
Again, on deuteron target  it measures the sum of the valence-quarks transverse distributions:
\be
h_{1Q_V}(\xb,\kt)&\equiv& h_{1u_V}(\xb,\kt)+h_{1d_V}(\xb,\kt)\\
h_{1Q_V}(\xb,\kt)&=&\frac{1}{2}\,{\cal
N}\T^{Q_V}(\xb)\,\left[Q_V(\xb)+\Delta Q_V(\xb)\right]\,\frac{e^{-\kt^2/\avk\T}}{\pi\avk\T},
\ee
where  $\avk\T$ and ${\cal N}\T^{Q_V}(\xb)$ are the unknown quantities.

Common for BM and Collins difference asymmetries is that they are convoluted with the same Collins FF, $\Delta^N D_{h^+/u_V\uparrow}(z_h,p_\perp)$, and
that the $\xb$ and $z_h$ dependencies factorize. This allows to relate the $z_h$-dependent BM and Collins asymmetries.
This relation is especially simple for small enough bins in $Q^2$ with a completely fixed $z_h$-dependence and  the same for all final
hadrons:
  \be
 \frac{A\BM^{h-\bar h} (z_h)}{A_{UT}^{Coll, h-\bar h}(z_h)}=
  \frac{\lambda_{Q_V}}{2}\,
  \frac{z_h\,\sqrt{\avPT_{\widetilde T}}}{\avPT_{\widetilde{BM}}}\,
  \bar B
  ,\quad h=\pi^+,K^+\label{BM-Coll}
  \ee
where $\avPT_{\widetilde{BM}}$ and $\avPT_{\widetilde T}$ are  fixed
by the assumed Gaussian form:
 \be
 \avPT_{\widetilde{BM}} = \avp\C +z_h^2\avk\S, \quad \avPT_{\widetilde T} = \avp\C +z_h^2\avk
\ee
 and
$\bar B$ is independent of, both, $z_h$ and the final hadron. Eq.
(\ref{BM-Coll}) will hold only if $\Delta^N D_{h^+/u_V\uparrow}$
enters both $A\BM^{h-\bar h}$ and $A^{Coll,h-\bar h}$ and thus,  it
 presents a remarkably simple test, involving only measurable quantities, not only of the standard parametrization, but of the whole QCD picture.
%%%%%%%%%%%%%%%%%%%%%%%%%%%%%%%%%%%%%%%%%%%%%%%%%%%%%%%%%%%%%%%%%%%%%%%%%%%%%%%%%%%%%%%%%%%%%%%%%%%%%%%%5
\section*{Acknowledgments}
E.C. is grateful to Oleg Teryaev for helpful discussions, to Bakur Parsamyan
for stressing the experimental importance of the proton data, and for the hospitality  of
 JINR  where the paper was finalized.
 E.L is grateful to The Leverhulme Trust for an Emeritus Fellowship.

%%%%%%%%%%%%%%%%%%%%%%%%%%%%%%%%%%%%%%%%%%%%%%%%%%%%%%%%%%%%%%%%%%%%%%%%%%%%%%%%%%%%%%%%%%%%%%%%%%%%%%%%%%%%%%%%%%%%%%%%%

\end{document}